# Computational modeling of the class I low-mass protostar Elias 29 applying optical constants of ices processed by high energy cosmic ray analogs

W. R. M. Rocha

Instituto de Pesquisa & Desenvolvimento, Universidade do Vale do Paraiba, Sao Jose dos Campos, SP 12244000

willrobson88@hotmail.com

and

S. Pilling

Instituto de Pesquisa & Desenvolvimento, Universidade do Vale do Paraiba, Sao Jose dos Campos, SP 12244000

sergiopilling@yahoo.com.br

Not to appear in Nonlearned J., 45.

## ABSTRACT

We present the study of the effects of high energy cosmic rays (CRs) over the astrophysical ices, observed toward the embedded class I protostar Elias 29, by using computational modeling and laboratory data. Its spectrum was observed with Infrared Space Observatory - ISO, covering 2.3 - 190 µm. The modeling employed the three-dimensional Monte Carlo radiative transfer code RADMC-3D (Dullemond 2012) and laboratory data of bombarded ice grains by CRs analogs, and unprocessed ices (not bombarded). We are assuming that Elias 29 has a self-irradiated disk with inclination i = $60°$, surrounded by an envelope with bipolar cavity. The results show that absorption features toward Elias 29, are better reproduced by assuming a combination between unprocessed astrophysical ices at low temperature ($H_2O$, CO, $CO_2$) and bombarded ices ($H_2O:CO_2$) by high energy CRs. Evidences of the ice processing around Elias 29 can be observed by the good fitting around 5.5-8.0 µm, by polar and apolar ice segregation in 15.15-15.25 µm, and by presence of the $CH_4$ and HCOOH ices. Given that non-nitrogen compounds were employed in this work, we assume that absorption around 5.5-8.0 µm should not be associated with $NH_4$ ion (Schutte & Khanna 2003), but more probably with aliphatic ethers (e.g. R1-$OCH_2$-R2), $CH_3CHO$ and related species. The results obtained in this paper are important, because they show that the environment around protostars is better modeled considering processed samples and, consequently, demonstrates the chemical evolution of the astrophysical ices.

Subject headings: protostellar disk, ISM molecules, radiative transfer, laboratory experiments, individual object: Elias 29

## 1. Introduction

The star formation process takes place inside molecular clouds, where dense cores collapse to form young stellar objects (YSOs), often called protostars. Following the initial contraction phase, these objects continue their physical evolution. Several observations and computational models showed that protostars can be classified in four-classes, Class 0 - III (Lada & Wilking 1984; Lada 1987), which suggests an evolutionary sequence among them. A Class I source, which is the subject of this paper, is a T Tauri star surrounded by infalling envelope, which in some situation has bipolar cavities. Scatter light models show that cavities are necessary to fit better the observed flux in the near-IR (Kenyon et al. 1993; Whitney et al. 1997). In addition to physical evolution of young stars, their environment evolves chemically, due to rich physicochemical processes which exist among gas, ices and high energy radiation.

In densest regions of the protostellar disk and envelope, the molecules in the gas-phase can condense onto dust grains by adsorption process, such as physisorption and chemisorption, forming ice mantles. Once formed, the astrophysical ices (as they are also called) can be subjected to FUV or X-ray radiation due to protostar, as well as bombardment by cosmic rays (CRs) from interstellar medium. Such interactions are important, because they can change ice mantle properties, and also return molecules (and new molecules) to the gas-phase by desorption processes (Tielens & Charnley 1997; van Dishoeck & Blake 1998). In other words, the ices experienced chemical changes by thermal or radiative processing. However, having average energies of several MeV, the CRs are able to penetrate deeply into the envelope or disk, compared to FUV or X-rays photons. Therefore, since the ices are formed in the densest parts of the disk or envelope, the chemical effects driven by FUV or X-rays photons on the ices, are very small, compared with CRs. Instead, such photons are very important to regulate the gas-phase chemistry, as for example, to dissociate molecules (HCN, $H_2O$, CO, $H_2$), or ionize He, considering X-rays (Henning & Semenov 2013).

Pilling et al. (2010a,b, 2012) have suggested that bombardment of ices by CRs analogs, may help to justify some features in the infrared spectra of YSOs. In addition, as discussed by Pilling et al. (2011), the chemistry of the ice depends on the temperature, besides the stellar radiation field (IRF) or interstellar radiation field (ISRF). A recent paper from Cleeves et al. (2014) highlights the importance of CRs for enhancement of the chemistry inside protoplanetary disks. From this manuscript, the authors shows that, if a model does not contain CRs as an ionizing factor, then the ionization level in the internal regions of the disk is too small to support a high rate of ion-molecule reactions.

The protostar Elias 29 (Elias 1978), also known as Elias 2-29 and WL 15 (Wilking & Lada 1983) is located in ρ Ophiuchi molecular cloud and is classified as class I YSO with an age range of 0.4 - 5 × $10^5$ yr (Chen et al. 1995; Boogert et al. 2000; Enoch et al. 2009). The coordinates of Elias 29 (J2000) are α = $16^h27^m09.3^s$ and δ = $-24°37'21''$. Despite several investigations, the distance and luminosity of this object given in the literature are still quite divergent: d ∼ 160 pc (Whittet 1974; Boogert et al. 2000; Boogert et al. 2002a), d ∼ 125 ± 25 pc (de Geus et al. 1989; Lommen et al. 2008), L = 36 $L_\odot$ (Chen et al. 1995; Boogert et al. 2000; Boogert et al. 2002a) and L = 13.6 $L_\odot$ (Evans et al. 2003; Lommen et al. 2008). Elias 29 is also surrounded by several other YSOs presents in the ρ Ophiuchi molecular cloud.

The inclination of Elias 29 has been estimated from spectroscopic features and more recently from images obtained by Hu´elamo et al. (2005). Kenyon et al. (1993) argue that deep absorption silicate bands, such as observed in Elias 29, are typical of disk with edge-on inclination. However, studies from Boogert et al. (2002a) shows that due to flatness of the Elias 29 spectrum, from 4 μm to 100 μm, the disk inclination must be less than $60°$.

From this result, the deep absorption features observed from spectral energy distribution (SED) due to ices and silicate, must originate in the envelope or foreground clouds, found in the same study. In this work, we adopt a fiducial inclination of i = 60°, as for example, Lommen et al. (2008) and Miotello et al. (2014). In the studies from (Hu´elamo et al. 2005), the authors obtained a direct image of Elias 29, using a polarimetric differential technique with a NAOS CONICA (NACO) instrument and adaptive optics facility from Very Large Telescope (VLT). The results show that a full edge-on inclination is improbable for Elias 29. Besides the presence of the disk, the authors also detected in near-IR (H and K band) a bipolar hydrogen molecular outflows in the NW-SE direction (assuming N to top and E to left) and a dark lane in the NE-SW direction. The presence of CO and molecular hydrogen was also observed in the outflows of Elias 29 by G´omez et al. (2003), and Zhang et al. (2013), respectively.

Öberg et al. (2011) and references therein show that most abundant molecular species around low and high mass young stars are $H_2O$, CO, $CO_2$, $CH_3OH$, $NH_3$, $CH_4$ and $OCN^-$. By using spectral data from Infrared Space Telescope (ISO), Boogert et al. (2000) has shown that inventory of ices around Elias 29, is composed mainly of $H_2O$, CO, and $CO_2$. However, several absorption features between 5.5 - 8.0 µm were also observed toward this source, including the prominent band at 6.85 µm. This last feature, has been often attributed to $NH_4^+$ ion (Demyk et al. 1998; Hudson et al. 2001; Schutte & Khanna 2003). Observe, however, that condensation of gas over dust grains, leads in many cases, to formation of dirty ices, such as $H_2O:CO_2$ (Öberg et al. 2011). Once that different molecular species are trapped in the same ice matrix, the chemistry induced by high energy radiation should be rich. Nevertheless, it is not conclusive from Boogert et al. (2002a), if the ices in the disk or envelope have experienced some interaction with high energy radiation, inducing chemical changes in their properties.

In this paper, we investigate if the ices observed toward Elias 29, show evidence of the thermal or radiative processing, due to interaction with high energy CRs coming from interstellar medium. For that, we present the better model for Elias 29 SED and the resolved simulated IR images (from near to far IR). The general parameters that constrain our modeling were obtained from observational data in the IR, taken from literature (Boogert et al. 2002a; Huélamo et al. 2005; Lommen et al. 2008; Beckford et al. 2008; Miotello et al. 2014). The model was built employing the radiative transfer code RADMC-3D (Dullemond 2012), assuming the presence of not bombarded ices (unprocessed ices) (Rocha & Pilling 2014) and bombarded ices by CR analogs (processed ices) (Pilling et al. 2010a), distributed by protostellar disk and envelope. By using the temperature of the disk and envelope calculated by RADMC-3D code, the position of the ices employed in the modeling were fixed, because the ices cannot occupy regions where the temperature is enough to return the frozen molecules to the gas-phase. For comparison, we also show the fit for the model using only unprocessed ices, which is the usual method observed in the literature (e.g. Boogert et al. (2002a); Pontoppidan et al. (2005)). However, such methodology cannot reproduce the chemical evolution features observed in Elias 29 spectrum. This paper shows that, a powerful radiative transfer code combined with processed ices to simulate the chemical evolution of the astrophysical ices, is fundamental to better characterize the protostellar environment. Such evidences of the chemical evolution of the ices around Elias 29, is given by presence of several absorption features between 5.5 - 8.0 μm, such as $CH_4$ and HCOOH, and by the polar and apolar ice segregation, seen by splitted profile in 15.15 and 15.25 μm. Furthermore, we showed that absorption bands between 5.5 - 8.0 μm are better related with CH bonds, than $NH_4$ ion. Additionally, in this work, we verified which size of dust grains better fits the observed spectrum of Elias 29 in the infrared.

This paper is structured as follow: In section 2 we present how the observation and laboratory data were obtained and how the modeling was employed. Section 3 shows the results of this paper and in the section 4, is summarized the predictions of the model for Elias 29.

## 2.  Methodology

### 2.1.  Observations

#### 2.1.1.  Elias 29 SED

The spectrum used in this paper for Elias 29, was obtained from the public database[1] of the Infrared Space Observatory (ISO) observed with two instruments: Short Wavelength Spectrometer (SWS - de Graauw et al.  (1996)) for 2.3 - 45 $\mu$m and Long Wavelength Spectrometer (LWS - Clegg et al.  (1996)) for 45 - 190 $\mu$m. The power resolutions R = $\lambda/\Delta\lambda$ are $R_{SWS}$ = 400 and $R_{LWS}$ = 200, respectively. The exposure time using each instrument on the target were 3454 s and 2611 s, respectively.

The spectrum was reduced automatically, using the Off-Line-Processing package (OLP). As informed in the on-line database, the two spectra have good quality, without dark current problems and are scientifically validated. Additional information about the reduction processing can be found in ISO Data Analysis Software[2] and Boogert et al. (2000). Figure 1a presents a field of view of 38 × 32 arcmin taken from MIPS camera in 24 $\mu$m from Spitzer Space Telescope. This image shows a very dense region with several embedded sources, where Elias 29 is indicated by a red arrow. Figure 1b shows the IR spectrum of Elias 29 obtained from ISO (SWS + LWS), dominated by absorption bands due to silicate and ices along the line of sight. In addition to strong bands observed in this spectrum, other weak features are also detected (Boogert et al. 2000).

---

[1]http://iso.esac.esa.int/

[2]http://iso.esac.esa.int/archive/software/

## 2.1.2. IRAC bands photometry

To determine the evolutionary stage of Elias 29, images were taken of the ρ Ophiuchi cloud from public archive of the IRAC/SPITZER camera[3] in four filters (3.6 μm, 4.5 μm, 5.8 μm and 8.0 μm). Photometric and instrumental corrections have previously been made (Fazio et al. 2003; Hora et al. 2008).

The astrometry and photometry of each image were performed by using the Starfinder code (Diolaiti et al. 2000), by PSF fitting. Such procedure allowed us to build a Color-Color diagram, and determine the evolutionary stage of Elias 29, as showed in Figure 2.

The boundary boxes were taken from (Megeath et al. 2004), to define the regions for YSOs class I and II. This diagram also shows that Elias 29 classification agrees with previous study using a BLT (Bolometric Luminosity and Temperature) diagram from (Chen et al. 1995); therefore, it can definitively be classified as class I protostar.

## 2.2. Opacity calculation

One of the most important parameters of radiative transfers models are the optical constants (refractive index) of the matter. However, physical properties such as the size and geometry of absorbers (grains) and its temperature are also essential. Such parameters together can be written as one simple parameter called opacity. Because opacity parameters are a key part of the radiative transfer calculation, we show in this paper how it was calculated, considering two models for the nature of the grain around Elias 29. In this work, we consider the label "Model 1" for opacities of only unprocessed ices and "Model 2" for more realistic ices that were previously processed by CR analogs (taken from Pilling et al. (2010a)), combined with unprocessed ices.

---

[3]http://sha.ipac.caltech.edu

In Model 1, we adopted the usual methodology in the literature (Pontoppidan et al. 2005), using only unprocessed ices composed of $H_2O$, $CO_2$, CO and silicate ($MgFeSiO_4$) combined with amorphous carbon. The more realistic approach is described by Model 2, supposing that a fraction of the material, was processed by energetic CRs coming from the interstellar medium. In this case, Model 2 takes into account a processed mixture of $H_2O:CO_2$ (1:1) (Pilling et al. 2010a) in addition to pure grains of $H_2O$, $CO_2$, CO and silicate. Previous papers have usually used opacities of coagulated grains from (Ossenkopf & Henning 1994) and (Weingartner & Draine 2001) to simulate the presence of the grains and ices in the interstellar medium. In this paper, such data cannot be applied because we are using processed ices, which are not available in the literature.

### 2.2.1. Laboratory experiments

In the interstellar medium, frozen molecules on the dust grains are exposed to ionizing radiation which trigger physicochemical processes, allowing chemical changes. Such processes can be reproduced in the laboratory, leading to the understanding of how the chemistry in the star-forming regions and their environment evolves.

A typical set-up of the experiment is presented in Figure 3. Briefly, the samples were deposited onto substrates (e.g. CsI, ZnSe) previously cooled to cryogenic temperatures ($\sim$ 10 K), in the high vacuum regime ($p < 10^8$ mbar). Using this procedure, the molecules from gas or liquid are condensed to represent the astrophysical ices. During the experiments, these ices can be bombarded by CRs analogs or photons to simulate different input energies coming from interstellar medium or young star. Fourier Transform InfraRed spectroscopy (FTIR) was used a diagnostic tool to detect the molecular absorption bands in the sample.

In this paper, the astrophysical ices used in Model 1 and 2 were simulated in the laboratory to obtain the absorbance data in the infrared. The experimental procedure employed for non-bombarded ices can be found in Rocha & Pilling (2014), and for the processed ice at work of Pilling et al. (2010a). The processed ices employed in this work (Model 2) was obtained from the mixture of $H_2O$ and $CO_2$, bombarded with heavy ions ($^{58}Ni^{13+}$ ion projectiles with energies of 52 MeV) up to a final fluence equal to $1 \times 10^{13}$ ions/cm$^2$ (Pilling et al. 2010a), to simulate the effects of the high energy CRs. At the end of the experiment, the formation of new infrared bands associate with new molecular species were observed.

Details about the chemistry behind the ice processing by CRs, can be found in Bennett et al. (2005) and Pilling et al. (2010a,b, 2011). In summary, if CRs particles penetrate into ice-forming regions, they lose their energy almost exclusively due to electronic interaction with the target molecules, allowing temperature increase among other effects. Such energy exchange may be enough to break chemical bonds of the molecules in the solid phase, allowing the formation of new molecular species.

### 2.2.2. Dust model

Because it is necessary to model the ices and silicate abundances independently for Elias 29, the opacities were calculated for each species, using as input data, the results from experiments in the laboratory and the optical constants. The first step was to calculate the optical constants of the ices, given by equation $m = n + ik$, where the $n$ value represents the real part and $k$ values the complex part, and both are function of the wavelength. They were calculated directly from absorbance data in the infrared obtained from experimental works. This procedure was performed with the NKABS code, which is a free code developed in Python Programming Language (Rocha & Pilling 2014). This code is based on Lambert-Beer law, Kramers-Kronig relationship, and Mclaurin's methodology. Figure 4 presents the graphics for the optical constants of grain employed in this paper. The values for CO ice and silicate and amorphous carbon dust were taken from online database[4], [5], [6].

After the calculation of optical constants (refractive index in the infrared), the absorption and scattering opacities of the dust and ices were calculated employing the Mie theory. To proceed with the calculations, we are supposing that the grains are spherical, with sizes defined by a MRN distribution, given by a power law: $dN(a)/da \propto a^{-3/5}$, where $a_{min}$ and $a_{max}$ defines the boundary size of the grains Mathis et al. (1977). Results from Weingartner & Draine (2001) indicate that in the interstellar medium, the dust grains are composed of silicates species and amorphous carbon. The size of such dust grain, varies from $a_{min} = 0.005$ - $a_{max} = 0.25$ µm. In addition, inside dense clouds or in the star-forming regions, the grains can be mixed due to physical process, such as turbulence. This assumption is adopted in this paper, as well as in previous papers in the literature (Ossenkopf & Henning 1994; Weingartner & Draine 2001; Pontoppidan et al. 2005). Additionally, there are observational evidences of the presence of large grains (a ∼ 1 µm) inside of dense clouds in the interstellar medium (Steinacker et al. 2011). In the current work, the presence of silicate grains with different sizes around Elias 29 were investigated in our simulations: (i) $a_{grain} = 0.005$ - $0.25$ µm, (ii) $a_{grain} = 0.025$ - $0.70$ µm from Beckford et al. (2008), and (iii) $a_{grain}$ = 0.25 - 1.0 µm. The size range of the employed ice grains in this work were considered to be: $a_{ice}$ = 0.0125 – 0.125 µm. Figure 5 presents the opacities calculated for grains used in this paper, considering the case (ii) for dust grains.

[4]http://www.strw.leidenuniv.nl/lab/databases/isodb/isodb.html

[5]http://www.astro.uni-jena.de/Laboratory/OCDB/data/silicate/amorph/olmg50.lnk

[6]http://www.astro.uni-jena.de/Laboratory/OCDB/data/carbon/cel800.lnk

## 2.3. Modeling Elias 29

A lot of parameters are necessary to model a young star, which means that high degeneracy is involved in the calculations. Previous papers from Robitaille et al. (2007); Gramajo et al. (2010); Whitney et al. (2013) show that is important to define a range of values for each parameter involved in the modeling process, and after that, try to find the best set of them. For Elias 29, such parameters range were constrained from observational results published in previous papers from Boogert et al. (2002a); Huélamo et al. (2005); Lommen et al. (2008); Beckford et al. (2008); Miotello et al. (2014) and can be seen in Table 1.

To model the observed spectrum of Elias 29 and its images in the IR, a three-dimensional Monte Carlo radiative transfer code RADMC-3D (Dullemond 2012) was used, with the opacities described in the previous section. In this code, we assume that density structure for Elias 29 is axisymmetric, although the photons are able to interact in the three dimensions. This approximation makes the problem essentially 2-D with coordinates (r, θ), and thus, much less computationally heavy. The Monte Carlo procedure to calculate the temperature of the system is describe in (Bjorkman & Wood 2001), and its result is used to determine the position of the ices, where they can survive in the solid phase as well as to create the SEDs and images of Elias 29 in specific wavelengths. During the simulation the Monte Carlo calculation incorporated absorption and isotropic scattering, and was considered the size aperture of $100''$ to build the SEDs. Such aperture is necessary to compute the flux of entire system (disk and envelope), as will show in the next sections.

Figure 6 shows a schematic diagram of Elias 29 environment, together with the ionizing radiation due to protostar and cosmic rays. This paper is focusing on the effects of CRs on ices observed toward Elias 29. This figure also shows the inclination angle of observation and the chemical scenarios seen toward Elias 29.

### 2.3.1.   Central object

Because Elias 29 is a protostar heavily embedded, the high obscuration prevents optical spectroscopic measurements of effective temperature $T_{eff}$. Thus, for this kind of protostars, Myers & Ladd  (1993) proposed to use the bolometric temperature $T_{bol}$ to characterize them. The values estimated for Elias 29 in the literature are 410 K (Chen et al.  1995) and 390 K (Evans et al.  2003). Nevertheless, in this paper, an estimation of effective temperature was used for Elias 29 protostar based on McClure et al.  (2010), that employed the extinction law toward ρ Ophiuchi cloud to determine the spectral type of several YSOs and Miotello et al.  (2014), that obtained data in longer wavelengths. From these observational constraints, the physical properties should be $T_{eff}$ = 4913 K and R = 5.8 $R_\odot$, meaning that L = 17.5 $L_\odot$ and $T_{eff}$ = 4786 K and R = 5.9 $R_\odot$, meaning that L = 16.3 $L_\odot$, respectively. In this paper, the considered ranges were R = 5 - 12 $R_\odot$ and T = 4000 - 5000 K. From our modeling, the better values were $T_{eff}$ = 4880 K and R = 5.7 $R_\odot$ and L = 16.5 $L_\odot$.

### 2.3.2.   Disk and envelope geometry

The images from Hu′elamo et al.  (2005) constrain the flared disk up to around 175 AU. However, is important to note that such disk may have a non-flared (self shadowed) component at large radii that cannot be seen by in the scattered light in the near-IR.

Therefore, the images in the near infrared constrain a lower limit of the disk size, if a non-flared part is present.

The Elias 29 disk was modeled by using a density structure defined by power law along the radius and a Gaussian function in height scale, given by

$$\rho_{disk}(r,\theta) = \frac{\Sigma_0 (r/R_0)^{-1}}{\sqrt{2\pi} H(r)} exp \left\{ -\frac{1}{2} \left[ \frac{r\cos\theta}{H(r)} \right]^2 \right\} \qquad (1)$$

where $\theta$ is the angle from the axis of symmetry, $\Sigma_0$ is the surface density at outer radius $R_0$, which is related with the disk mass, $M_{disk}$ and $H(r)$ is the disk scale height. It is given by $H(r) = r \cdot H_0/R_0 \cdot (r/R_0)^{2/7}$, defined to the self-irradiated passive disk proposed by Chiang & Goldreich (1997).

The inner radius of the disk was calculated based on the sublimation temperature of the silicate, using $R_{in} = R_*(T_*/T_{in})^2$ from Dullemond et al. (2010). Assuming the silicate grains sublimate at $T_{in} = 1500$ K, then $R_{in} = 0.36$ AU for Elias 29. Following Lommen et al. (2008) and Miotello et al. (2014), the outer radius of the disk, $R_{out} = 200$ AU and the ratio $H_0/R_0 = 0.17$ were adopted in this paper.

Besides the disk, Elias 29 is surrounded by a large envelope, given the radial intensity profile obtained from the 1.3 mm map (Motte et al. 1998), modeled by Boogert et al. (2002a). Such profile, is better fitted by the presence of the envelope extended up to 6000 AU, in adition to the disk. To describe it, we are considering a static envelope, whose density is given by a following power-law:

$$\rho_{env}(r) = \rho_0 \left( \frac{R_{out}}{r} \right)^{1.5} \qquad (2)$$

where $\rho_0$ value is the density at the outer radius.

Given the observations of the jets in Elias 29 (Gómez et al. 2003; Zhang et al. 2013), is reasonable to consider a cavity in the envelope. The cavity aperture was constrained from near-IR images, combined with linear polarimetry technique toward ρ Ophiuchi, taken from Beckford et al. (2008). For Elias 29, the opening angle of the cavity estimated in the K band is $\theta = 30°$.

Table 1 summarizes the parameters used in Models 1 and 2, as well as the estimated range due to degeneracy involved in the modeling.

### 2.3.3. Grain spatial distribution

More realistic scenarios for modeling of protostellar disks and envelope should consider the sublimation temperature of the silicates and ices, as well as the possibility of the radiation and temperature processing them. Taking this into account, the silicate and ice grains should occupy specific regions around the protostar. The sublimation temperatures of the grains used in this paper can be found in the literature: (i) Silicate 1500 K (Gail 2010), (ii) $H_2O$ - 150 K, $CO_2$ - 75 K, $CO$ - 20 K (Collings et al. 2004) and (iii) $H_2O:CO_2$ bombarded, assumed to be 150 K.

Note, however, that frozen molecules trapped in $H_2O$ ice, will be desorbed at two different temperatures: (i) desorption of the species from the surface of the $H_2O$ ice and (ii) desorption from molecules trapped inside the bulk of the $H_2O$ ice (Collings et al. 2004). This means that volatile molecules can be placed at inner regions of the protostellar disk (Visser et al. 2009) and, consequently, more subjected to radiative and thermal processing.

In this paper, the dust grains were used in the spatial grids, and the ices, were employed where the temperature was lower than their sublimation temperature. This procedure was performed, by calculating the temperature of the structure by Monte Carlo methodology, firstly employing only the ice opacities. Next, was used this result to determine the positions of each ice around Elias 29, and a new temperature was calculated, to build the SED and images in the IR. Such procedure showed that pure CO ice cannot be present in the envelope along the line of sight of Elias 29, due to low sublimation temperature, and would not be seen in the modeled spectrum. To reproduce the CO observed in Figure 1, we are assuming that all CO is trapped in $H_2O$ ice. Such assumption that all species are constrained to disk and envelope, is not very realistic, because a fraction of the ice should be placed in foreground clouds.

## 3. Results

### 3.1. Spectral energy distribution

Figure 7 shows the spectral energy distribution for Model 1, employing only non-bombarded ices, similar to the one performed by Pontoppidan et al. (2005). Note however, that such modeling is able to reproduce a lot physical parameters of the protostar, but it was not able to reproduce the chemical changes induced by radiation (driven by CRs or high energy radiation) on the ice mantles, observed at wavelength around 5.5-8.0 μm and 15.15-15.25 μm. This means that, despite the strong bands observed in the Elias 29 SED (e.g $H_2O$, $CO_2$, CO, silicate) that were well fit by the modeling, weak features formed from ice chemistry cannot be probed by Model 1. On the other hand, the better model is showed in Figure 8, where the physical and chemical properties were reproduced. To reach this result, the presence of the bombarded ice ($H_2O:CO_2$) was fundamental to improve the chemistry of the modeling. Such processed ices was employed in this work in the Model 2. In both models, the average abundances of the ices relative with $H_2$ and dust mass of the disk and envelope were constrained from Boogert et al. (2000); Lommen et al. (2008). The section 3.3 shows how the abundances were derived in this paper.

Figure 7a and 8a show Elias 29 SED modeled with RADMC-3D covering 0.1 - 80000

µm. This graphic also shows the (i) observational spectrum from ISO (black line), (ii) blackbody of the protostar without extinction (dotted black line), (iii) cold foreground clouds (blue and green dashed lines), as well as the (iv) photometric data from bands 2MASS (1.23 µm, 1.66 µm, 2.16 µm), WISE (3.35 µm, 4.6 µm, 11.56 µm, 22.08 µm), SPITZER (8 µm), IRAS (12 µm, 25 µm, 60 µm, 100 µm) and in the longer wavelengths (1.1 mm, 3 mm, 3 cm, 6 cm). The photometric data in the IR regime were taken from Infrared Science Archive[7] and in the long wavelength regime from Lommen et al. (2008) and Miotello et al. (2014). Due to fact that Elias 29 protostar is heavily embedded in its birthplace, we should consider the presence of the cold foreground (12 - 16 K) clouds (Boogert et al. 2002) to better simulate the observed spectrum after 80 µm. The foreground emission of cold clouds was added to the SED by hand, and is not part of the radiative transfer simulation. These clouds produce a **modified** blackbody emission at long wavelengths (Hildebrand 1983; Ward-Thompson et al. 2000), sometimes called as graybody emission, and given by

$$S_\nu = \Delta\Omega B_\nu(T) \left[ 1 - e^{-\tau_\nu} \right] \tag{3}$$

where $\Delta\Omega$ is the solid angle of the source, $B_\nu(T)$ is the blackbody emission at temperature $T$. The optical depth is given by $\tau_\nu = k_\nu N_d$, where $k_\nu$ is the absorption mass **coefficient** given by $k_\nu = k_0(\nu/\nu_0)^\beta$, where $\beta$ is the opacity index, and $N_d$ the mass column density of the dust toward the line of sight. Adopting the $k_0$ at 1.3 mm as $k_{1.3}$ =0.01 (taken from Figure 5, assuming the dust-to-gas ratio of 0.01), was performed the fit by using two graybody to reproduce the Elias 29 SED at long wavelengths. The Table 2 shows the better parameters obtained with the fit.

Figure 7a and 8a, also show that near-IR fluxes are over strong extinction by dust. From the **difference** between the blackbody of the protostar without extinction (dotted line) and the modeled blackbody (red line), the extinction calculated for J band at 1.23 μm is $A_J$ = 15.3 mag. Assuming that visual extinction $A_V$ is related with $A_J$ by using $A_J$ = 0.282$A_V$ from Cambrésy et al. (2002), then the visual extinction toward Elias 29 is $A_V$ = 54.3 mag. Given the relation between visual extinction and column density of hydrogen from Bohlin et al. (1978), thus $N_H$ = 1.3 × $10^{23}$ cm$^{-2}$ toward line of sight of Elias 29. This value agrees with estimative presented in Boogert et al. (2000), calculated in $N_H$ = 0.5 - 1.2 × $10^{23}$ cm$^{-2}$. From McClure et al. (2010), it is possible calculate that $N_H$ = 0.8 × $10^{23}$ cm$^{-2}$ for Elias 29.

---

[7]http://irsa.ipac.caltech.edu

Figure 7b and 8b show, with detail, the fitting (red line), covering 2.3 - 200 μm over the observational data from ISO (black line). These panels highlight the good fit of the continuum from near-IR to far-IR. Figure 7c and 8c present strong absorption bands of $H_2O$ (3.07 μm), $CO_2$ (2.69 μm and 4.26 μm) and CO (4.67 μm) ices. As discussed by Rocha & Pilling (2014) the observed absorption profile of $H_2O$ indicate clearly its amorphous structure. Laboratory studies from Fraser et al. (2001) shows that $H_2O$ ice change from amorphous to crystalline phase at temperature around 120 K. This means that the temperature where $H_2O$ ice is condensed, must be lower than 120 K. Figure 7c and 8c also show a small shift of the modeled band of $CO_2$ and CO, compared with the observation. This happens because we are assuming that ice grains in the modeling, has a spherical geometry, as proposed by Mie theory. Studies from Boogert et al. (2002b) show that if are assumed ellipsoidal grains, instead spherical one, such bands can be better fitted.

Figure 7d and 8d, show that the silicate is the most strong band in this range (9.8 μm and 18 μm). Furthermore, ice grains containing molecules such as $H_2O$, $CO_2$, HCOOH and $CH_4$, observed toward Elias 29 (Boogert et al. 2000), also present vibrational modes in this part of the infrared spectrum. However, Figure 7d and 8d show that if only unprocessed ices are employed in the modeling (Model 1), the absorption bands between 5.5 - 8.0 are poorly fitted. Otherwise, if a fraction of bombarded ice ($H_2O$:$CO_2$) is added to unprocessed one (Model 2), the same range in the observed infrared spectrum of Elias 29 is better simulated. The inset graphic in Figure 8d, shows a detailed zoom of the fitting between 5.5 - 8.0 μm. Model 2 is also able to better fit the $CO_2$ absorption band at 15.15 μm and 15.25 μm, compared with Model 1.

Figure 9a presents, with details, the infrared spectrum of Elias 29, as well as the two Models developed by this work in the spectral region between 5.5 - 8.0 μm. Several molecules has vibrational modes in this spectral region and, therefore, the laboratory experiments are very helpful to identify the presence of specific molecules. Firstly, is observed the presence of a HCOOH ice, due to vibrational modes at 5.85 μm (C=H stretch vibrational mode) and at 7.25 μm due to CH deformation. Next, a deformation mode of $CH_4$ ice is also observed at 7.69 μm. Additionally, some complex molecules may be associated with the absorption bands at 6.85 μm and 7.01 μm. The band at 7.01 μm can be associated with the acetaldehyde molecule ($CH_3CHO$) due to $CH_3$ deformation mode. Studies from Bennett et al. (2005) shows that this molecule can be formed inside of dense molecular clouds when the astrophysical ice interacts with CRs. On the other hand, the nature of the band at 6.85 μm has been unclear. Usually in the literature, this band has been associated with the $NH_4^+$ ion from Schutte & Khanna (2003). However, in this paper we have shown that this feature can arise from a mixture containing only $H_2O$:$CO_2$ bombarded by CR analogs. In the current methodology, the molecular composition of the employed mixture, does not contains nitrogen atoms, and therefore, the feature at 6.85 μm is reproduced without the need for the $NH_4^+$ (ammonium) ion.

Additional evidence of the thermal processing of the ice inventory around Elias 29, can be observed from Figure 9b. The $CO_2$ ice, pure and partially crystalline at low temperature ($\sim$ 10 K) is split into two narrow bands at 15.15 μm and 15.25 μm. On the other hand, if the $CO_2$ ice is trapped in a $H_2O$ ice, forming a mixture ($H_2O:CO_2$) at low temperature, such absorption band presents a unique peak. However, laboratory experiments from Ehrenfreund et al. (1997) has shown that if a mixture of $H_2O:CO_2$ is submitted to the increasing of the temperature, the split profile of $CO_2$ at 15.15 μm and 15.25 μm arises again. Such behavior has been associated with the ice segregation between apolar ($CO_2$) and polar ($H_2O$) molecules in the ice, which begins around 40 K. The second peak increases together with the increasing of the temperature. Figure 9b shows that if is assumed pure crystalline $CO_2$ ice (Model 1), the band in 15.15-15.25 is poorly fitted. The better fit is reached if a thermally processed ice is present around Elias 29 (Model 2).

Figure 10 show that the regime between 5.5 - 8.0 μm can be associated with several another species that do not contain nitrogen atoms, as for example, aliphatic ethers (e.g. R1-OCH$_2$-R2) and related species. The problem of the complex molecules associated with this band, is the missing of strong absorption features in other wavelengths. Pilling et al. (2010b), show that if an ice mixture containing $NH_3$ is bombarded by CR analogs, the absorption band at 6.85 μm also arises. However, the $OCN^-$ is formed from ice bombardment as well, like Schutte & Khanna (2003) experiment. Perhaps, this means that the band at 6.85 μm is associated with $NH_4^+$ ion when $OCN^-$ is also formed. Otherwise, such band should be associated with vibrational modes of C-H bonds. We conclude that Model 2 produces a better fit than Model 1, and in Table 3 compare the observed and modeled SEDs (Fig. 8) at the central wavelengths of the ice absorption bands.

## 3.2.  Alternative models for grain size distribution

In this section, we show how the modeled SED of Elias 29 changes, by adopting three regimes of dust grain sizes (silicate and amorphous carbon) and following the MRN size distribution (see section 2.2.2 and Figure 5). It is importante to observe that the size distribution of the ice remains the same and the adopted parameters are the same of Model 2. Figure 11a shows the Elias 29 SED between 2.5 - 30.0 μm, where the blue dashed line is the fitting employing the small grain distribution (0.005 - 0.25 μm) and the green dashed line is the fitting employing large dust grains (0.25 - 1.0 μm). The better fit from near to far-infrared is given by size dust grains between 0.25 - 0.70 μm. The other sizes can fits the observational data besides 8 μm, but are not able to model the spectrum at near-IR. The infrared region between 2.5-4.0 μm for different size of dust grains is shown in Figure 11b, and highlights the fitting in the near-IR. The extinction caused by small or large dust grains cannot reproduce the observational SED in all wavelengths. On the other hand, dust grains with intermediary size (0.25 - 0.70 μm), are able to model with accuracy the spectrum of Elias 29. Such result is interesting, because dust grains with size of 0.6 μm were proposed by Boogert et al. (2000) to fit the Elias 29 SED, and with size of 0.7 μm were observed toward Elias 29 by Beckford et al. (2008).

Indeed, intermediary size of dust grains, following the MRN distribution, were able to model the Elias 29 SED in this paper. However, some caveats should be mentioned about this section. Miotello et al. (2014) show that dust grains with millimeter size can be present in the Elias 29 envelope, because they are necessary to fit the data at 1.1 mm and 3 mm from Submillimeter array (SMA) and Australian Telescope Compact Array (ATCA), respectively. However by assuming the MRN distribution in this paper, large grains are not able to reproduce the Elias 29 SED. Perhaps, a way to address this question would be a detailed study about the power-law of the grains size distribution focusing on Elias 29. An additional caveat about this section is that we are considering that ice opacities are not affected by changing the dust grain size. Moreover, we are supposing that the size of the ices follows strictly a MRN distribution given by small grains.

### 3.3. Grains abundances

To determine the average abundances relative to $H_2$, we employed the optical depth analysis explained in Boogert et al. (2000), because it allows to calculate the abundances of numerous ice species that are formed with the ice bombardment. Briefly, is defined the continuum curve over the observational SED, as simulated by the blackbody fits. Next, the equation $\tau(\lambda) = \ln(I_0(\lambda)/I(\lambda))$ is used to determine the optical depth $\tau(\lambda)$ in each wavelength, where the observed spectrum and the adopted continuum are given by $I(\lambda)$ and $I_0(\lambda)$, respectively. Thereafter, knowing the optical depth and the band strength of each absorption, the column density is calculated. The column density for hydrogen calculated in this paper was $N_H = 1.3 \times 10^{23}$ cm$^{-2}$ (see section 3.1). Lastly, the abundance of ices is given by ratio between the column density of ices and $H_2$. However, differently of Boogert et al. (2000), such methodology was applied by using the better model (Model 2), showed in Figure 8. This methodology allows to calculate the column density of the ices originated from the processing by CRs, but without the noise present in the observed spectrum. Table 4 shows the values obtained from this paper for Elias 29.

Figure 12a show the average abundances of ices calculated from Model 2, compared with values from Boogert et al. (2000) and Lommen et al. (2008). Because, we are assuming the presence of bombarded $H_2O:CO_2$ ice, the abundances of the processed material, was also calculated. The necessary amount of processed $H_2O$ and $CO_2$ ices to fit the observable spectrum was roughly 14 % and 7 %, respectively, relative to total abundance employed in the modeling. The limits greater than observed by Boogert et al. (2000) are due to bandwidth obtained from Model 2. The $CH_3OH$ and $CH_3CHO$ are not shown in this figure because the presence of such molecules is uncertain in this paper. Figure 12b, shows the mass of the disk and envelope used in the Model 2, compared with values from Lommen et al. (2008). The upper limit proposed for Elias 29 is $M_{disk} \leq 0.007\ M_\odot$ and $M_{envelope} \leq 0.058$ M$\odot$. The values calculated from Model 2 were $M_{disk} = 0.003$ M$\odot$ and $M_{envelope} = 0.028$ M$\odot$ and are showed in Figure 12b.

## 3.4. Density and temperature structure

Figure 13a and 13b present the density and temperature distribution for Model 2, emphasizing the disk and envelope, respectively. Such parameters were necessary to determine the regions where the ices employed in the Model 2 survive in the solid phase. The continuous lines in Figures 13a and 13b indicate the density contours for $H_2$ in cm$^{-3}$. Figure 13a shows that the disk density at the mid-plane reach values more than $10^7$ cm$^{-3}$, while in Figure 13b, the density of the envelope close to edge is equivalent to 140 cm$^{-3}$. The dashed lines represents the snow lines, which are limits that define regions around the protostar, where pure molecules survive in the solid phase. The inset panels in Figure 13a shows a detail of $CO_2$ and $H_2O$ snow lines.

The temperature map calculated from Model 2, using the RADMC-3D code, allows to conclude that water-rich-ices are present in the Elias 29 disk, just besides 8 AU, and in the envelope beyond 14 AU. Such difference is due to density distribution on the disk and envelope. On the other hand, the more volatile pure $CO_2$ ice, stay in the disk and envelope beyond 20 AU and 80 AU, respectively. However, pure CO ice just survive in the solid phase far away from radiation source. Due to disk inclination of i = 60$^\circ$ for Elias 29, the Figure 13b allows to conclude that most part of pure CO ice observed toward Elias 29, should not be placed in their envelope due to high temperature, but in cold foreground clouds. This confirms the observation of CO in the gas-phase toward Elias 29, as showed in Boogert et al. (2000). Although has been considered that all CO ice is trapped in the $H_2O$ ice, the most realistic scenario should take into account the radiative transfer in the foreground clouds containing pure CO ice.

There is evidences in the literature that snow lines moves away along the evolution of protostars from class I to photosphere objects. Zhang et al. (2013) report different water abundances in the T W HYA protoplanetary disk. The authors observed that between 0.5 - 4 AU, the abundance was calculated in $10^{-6}$ per $H_2$, while after 4 AU, was calculated in $10^{-4}$ per $H_2$. This observational data means that with time in passive disks, like Elias 29, the inner disk becomes very dry, while is formed a significant ice reservoir at large radii of the disk.

### 3.5. Infrared Images

In this section we present the images in the IR, modeled with RADMC-3D, assuming a more realistic scenario (Model 2). The images were created by using $1 \times 10^9$ photons package to run the Monte Carlo simulation. Then, the average intensity over width $\Delta\lambda$ of each filter, was calculated.

The comparison between observed images of Elias 29 in the infrared and modeled images (Model 2) are shown in Fig 14. Figure 14a present the observed image in the K band taken from Hu´elamo et al. (2005) with field of view FOV = $1' \times 1'$. Figure 14b shows the simulated image in the same band and in the same FOV employing Model 2. The difference in the brightening between observed and modeled images is possibly explained by external illumination originating from nearby stars. Whitney et al. (2013) show that the external illumination can dominate if the central source luminosity is very low, and/or the outer radius of the envelope is very large. From Figure 13b, we can observe that, the large model envelope around Elias 29 is very cold near the edge and therefore exhibits low surface brightness in the near-infrared. On the other hand, the presence of external illumination around Elias 29 would increase the temperature of the envelope, and consequently, increase the near-IR surface brightness. Furthermore, Figure 14c and 14d emphasizes the disk structure of Elias 29 in a FOV = $1.5'' \times 1.5''$. The observed image was taken from Hu´elamo et al. (2005), by using polarimetric technique with NACO instrument. Such image represents the normalized K-band polarization. The dark lane is seen in the NE-SW direction, in addition to scattering of the radiation in the NW-SE. The modeled image present the extinction of the radiation due to disk, as well as the scattering. From this image, we can conclude which the dark structure seen from observation is due to surface of the disk.

Figures 15 and 16, show the resolved images, emphasizing the disk structure and the envelope in the IR, respectively, as well as, the intensity in logarithmic scale. All panels in the Figure 15 has a field of view of $3.3'' \times 3.3''$ or 400 AU $\times$ 400 AU, while in the Figure 16 illustrate a field of view of $100'' \times 100''$ or 12000 AU $\times$ 12000 AU, assuming a distance of 120 pc, obtained from SED modeling.

## 4. Conclusions

We present a realistic theoretical model of the spectrum and image of Elias 29 protostar, using an axisymmetric, Monte Carlo radiative transfer code. An important characteristic of this paper was the use of radiative and thermal processed ices by high energy CRs, unlike of previous works. The main conclusions obtained in this paper, are summarized as follow:

1. The better model obtained here for this YSO indicates that Elias 29 is better explained by a class I protostar in the late accretion phase. Furthermore, Elias 29 should be surrounded by a disk, with inclination close to 60 °, to reproduce the observed SED and image in near-IR. Moreover, the environment of Elias 29 is better reproduced by considering grains larger than observed in interstellar medium, but somewhat lower than observed in dense regions of molecular clouds. Furthermore, a large fraction of the $H_2O$-rich ices observed toward Elias 29 are likely found in the cooler envelope regions, outside the $H_2O$ snow-line. On the other hand, the envelope is mostly too warm to support CO ice, implying most of the CO ice must be placed in foreground clouds.

2. There are strong evidences that the ices around Elias 29 has been processed by high energy CRs and by temperature, which differs from the interpretation in Boogert et al. (2002a). The fingerprint of this processing can be found from presence of the HCOOH, $CH_3CHO$ and $CH_4$, due to interaction of the $H_2O$ and $CO_2$ ices with high energy cosmic rays. Moreover, the band at 6.85 µm observed in the Elias 29 spectrum, can be associated with vibrational modes of C-H bonds, instead $NH_4^+$, due to absence of absorption feature associated with $OCN^-$. Additionally, the ice segregation observed in 15.15 and 15.25 µm is another evidence that the thermal processing, induced by CRs.

3. The image modeled in the K-band presents a good agreement with the observed image from ISSAC telescope and with results from Hu´elamo et al. (2005) for FOV = $1' \times 1'$ and FOV = $1.5'' \times 1.5''$. The resolved images of the disk also present a dark lane in near-IR, as observed in Hu´elamo et al. (2005). Additionally, the images emphasizing the envelope, shows characteristics of bipolar cavity, as observed in Elias 29.

4. This paper shows the importance in combining the radiative transfer simulations with laboratory data of processed astrophysical ices to understanding the absorption features of spectra of YSOs. Such procedure highlights the importance of CRs to drive the formation of complex molecules around protostars.

We would like to thank the anonymous referee for a careful review of the paper that allows their improvement. The authors also acknowledges the Financial support from

---

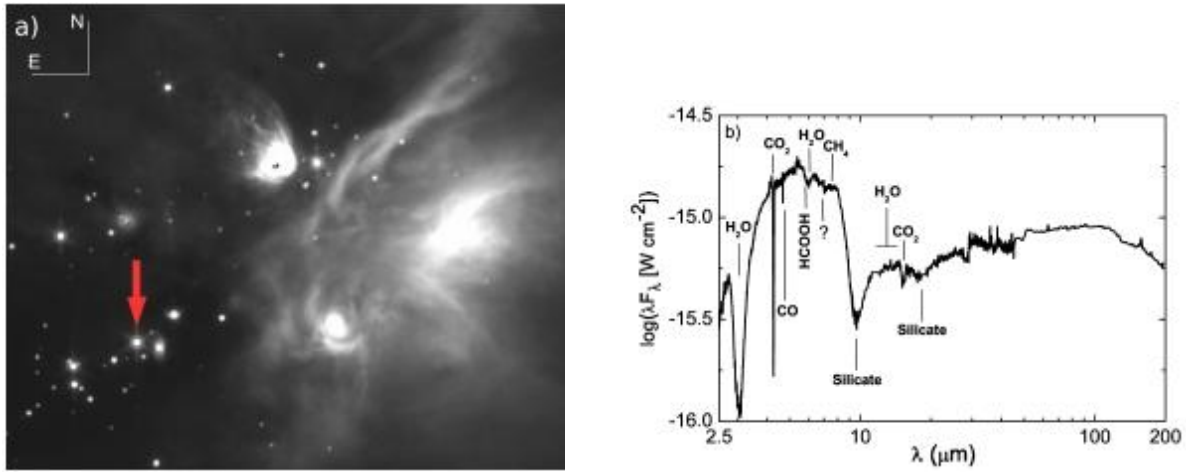

Fig. 1.— a) Rho ophiuchi molecular cloud region containing the low mass protostar Elias 29 (red arrow). b) IR spectrum of Elias 29 obtained from ISO observations (SWS + LWS). The strong absorption features due to silicate and ices are indicated. The question mark indicates the small absorption band in 6.85 μm (1459.8 cm$^{-1}$) which remains without conclusive assignment.

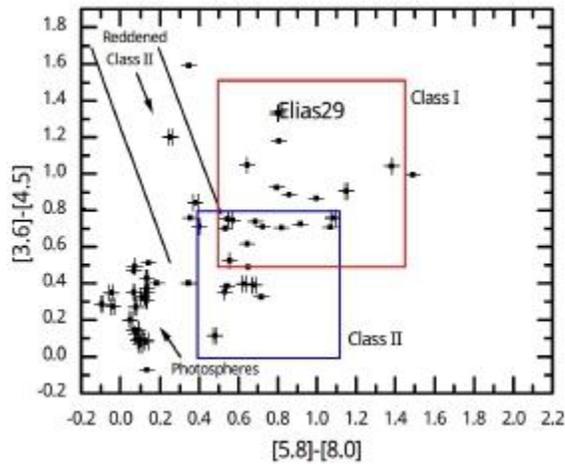

Fig. 2.— Color-Color Diagram for ρ Ophiuchi cloud. The boxes define regions for YSO class I and II. The black lines identify reddened class II protostars. The photospheres are main sequence stars, or YSO class III. Elias 29 is classified as Class I protostar and is indicated by its name.

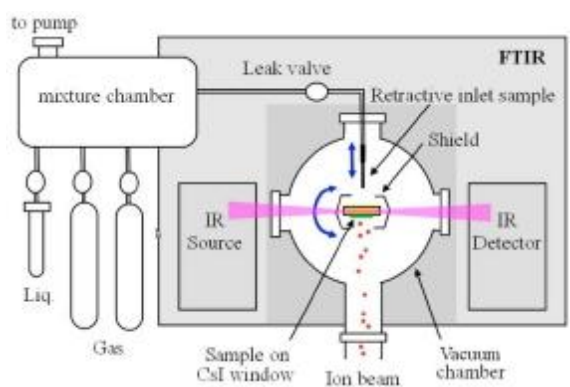

Fig. 3.— Schematic diagram of the experimental set-up. The ice is deposited on the substrate (e. g. ZnSe) where it is condensed due to lower temperature (generally ∼ 10 K). The infrared spectroscopy is used to detect the absorption bands before and after the irradiation of the ice with the ion beam. Extracted from Pilling et al. (2010a).

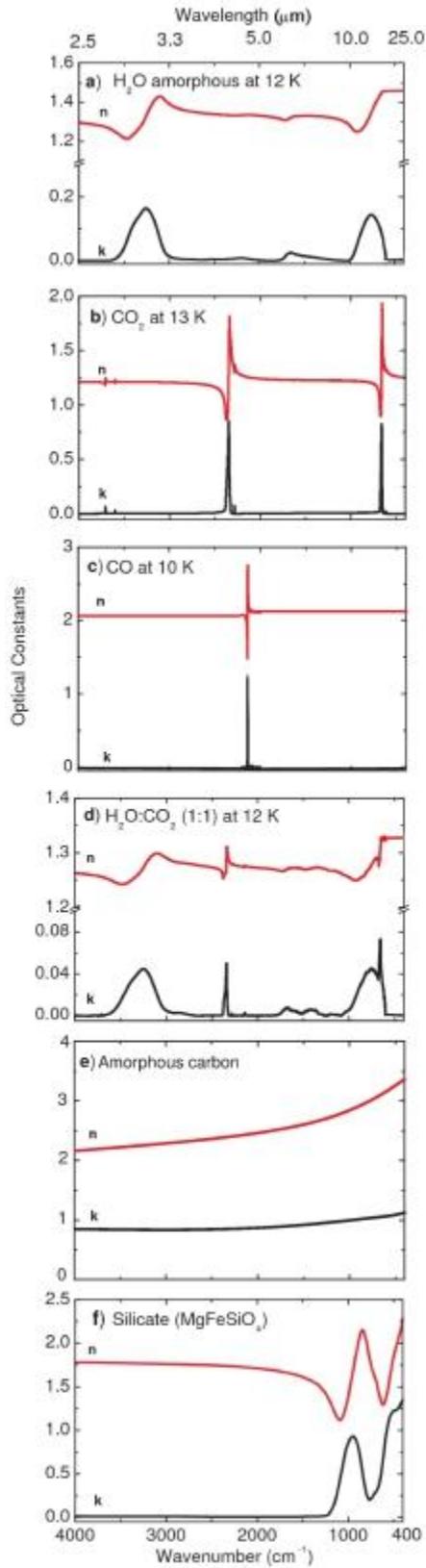

Fig. 4.— Optical constants n and k calculated using NKABS code, except for silicate and amorphous carbon

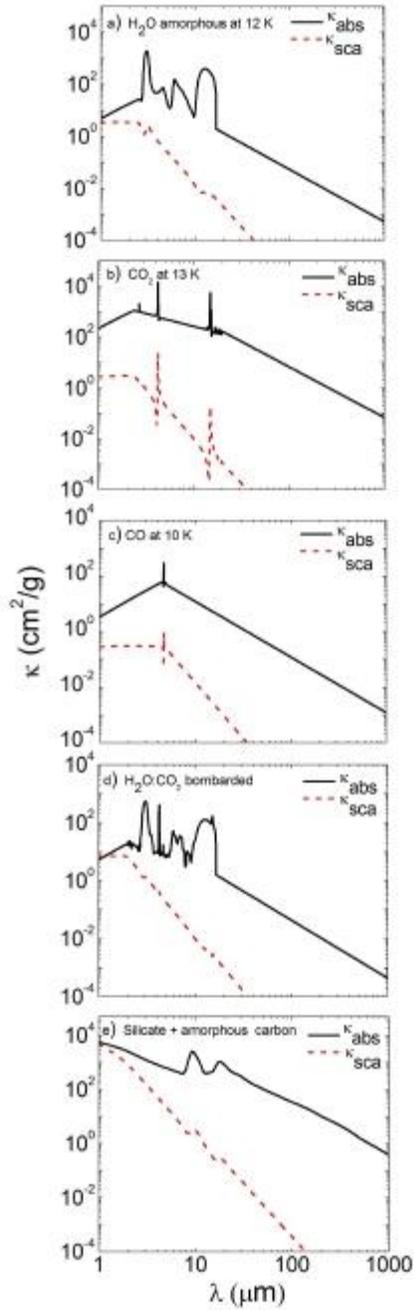

Fig. 5.— Opacities for ices and silicate with amorphous carbon. The continuous line represents the absorption opacity and the dashed line the scattering opacity.

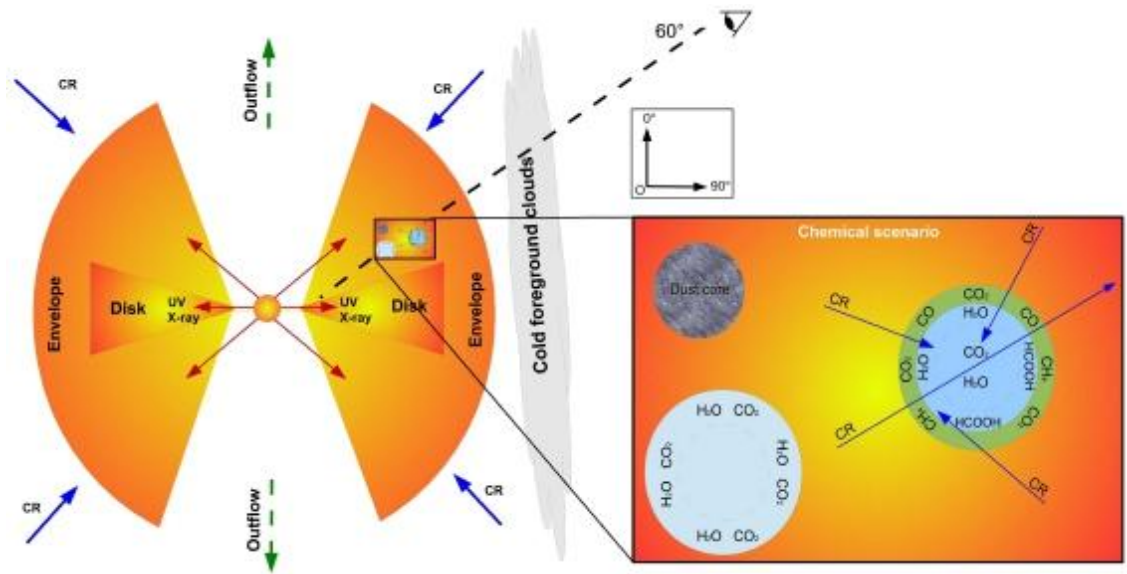

Fig. 6.— Schematic overview of the morphology and chemical scenario of Elias 29 employed in this paper. The line of sight crosses through of foreground clouds and envelope, indicating the angle i = 60°, taking as reference the angles showed in the inset panel.

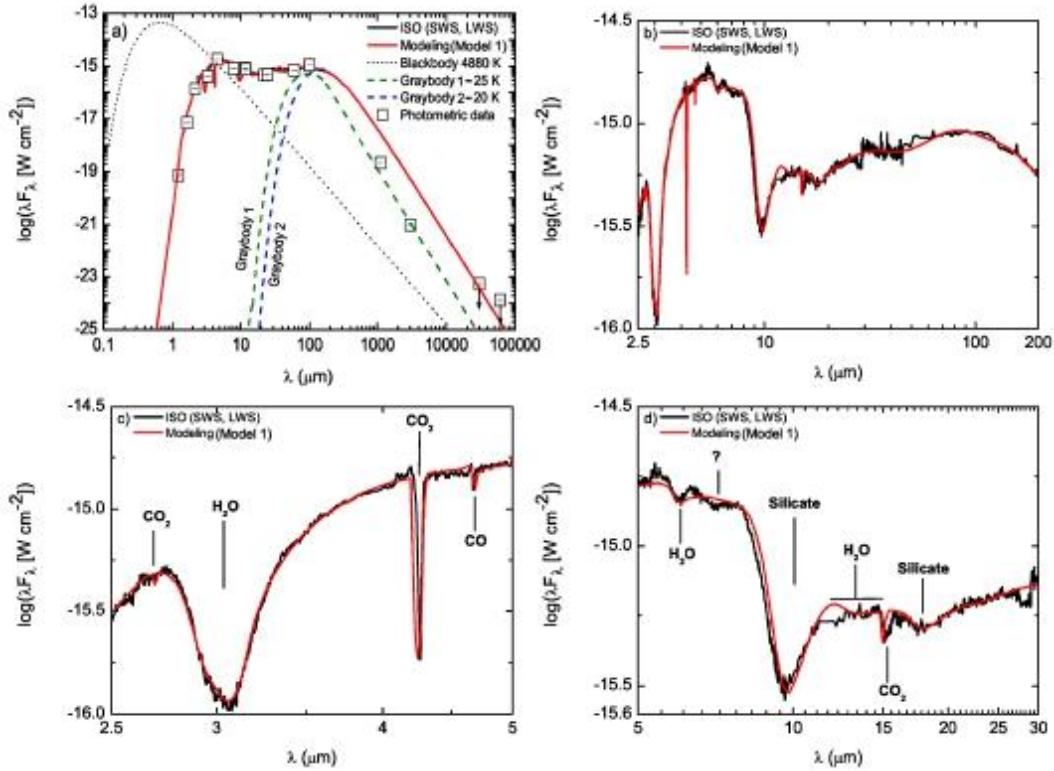

Fig. 7.— Modeling employing only unprocessed ices (Model 1) for Elias 29 SED. The four panels show the spectral energy distribution of Elias 29 (black line), modeled by RADMC-3D (red line). a) (i) the dotted black line represents the blackbody associated with 4880 K, (ii) blue and green dashed lines are the graybody associated with the cold foreground clouds (20 K and 25 K, respectively), (iii) photometric measurements from near-IR to sub-mm taken from public archives. b) Detail of panel a, covering from 2.5 - 200 μm. c) Detail of panel b, covering from 2.5 - 5.0 μm. This panel also shows the absorption features of CO, $CO_2$, and $H_2O$. d) Detail of panel b, covering from 5.0 - 30.0 μm. This panel shows the presence of absorption bands of $CO_2$, $H_2O$, and silicate.

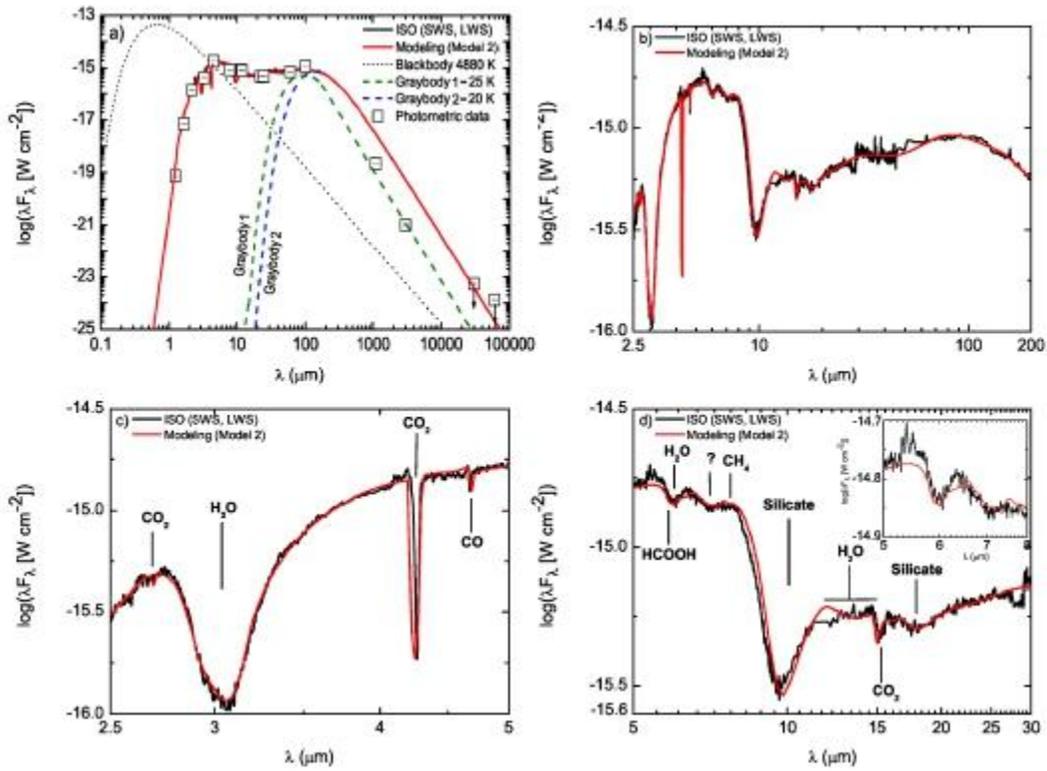

Fig. 8.— Better modeling employing processed ices (Model 2) for the Elias 29 SED. The lines and symbols are same of Figure 7. However, in the panel d, the absorption range from 5.5-8.0 μm was better fitted by Model 2, as well is indicated the presence of HCOOH and CH$_4$. The inset panel, is a zoom in on the region of the fitting in such region.

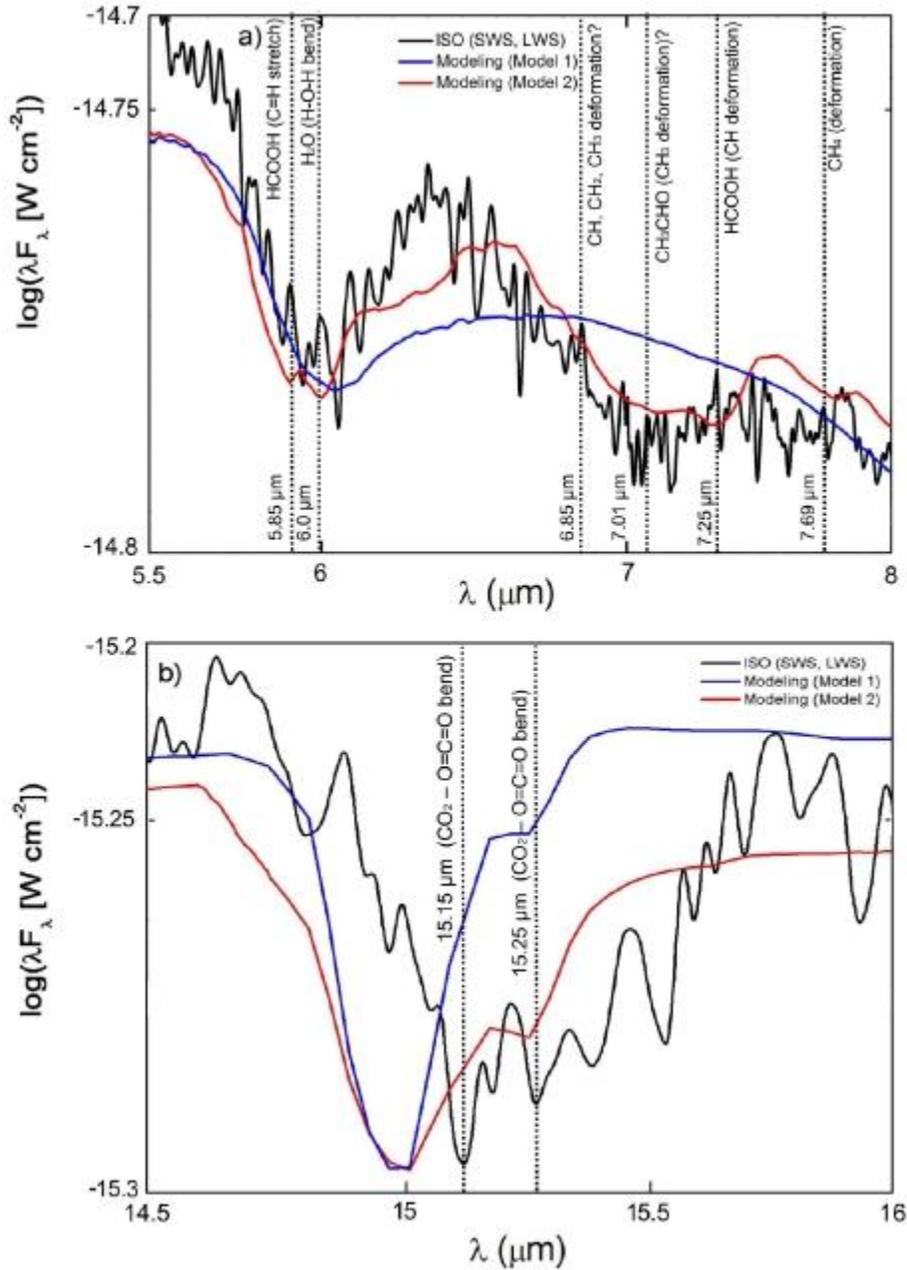

Fig. 9.— Evidences of the ice processing around Elias 29. Panel a shows the presence of HCOOH and CH₄. Furthermore, the range from 5.5-8.0 may be associated with C-H bonds, including CH₃CHO at 7.01 μm. Panel b shows evidences of the ice segregation in polar and apolar species.

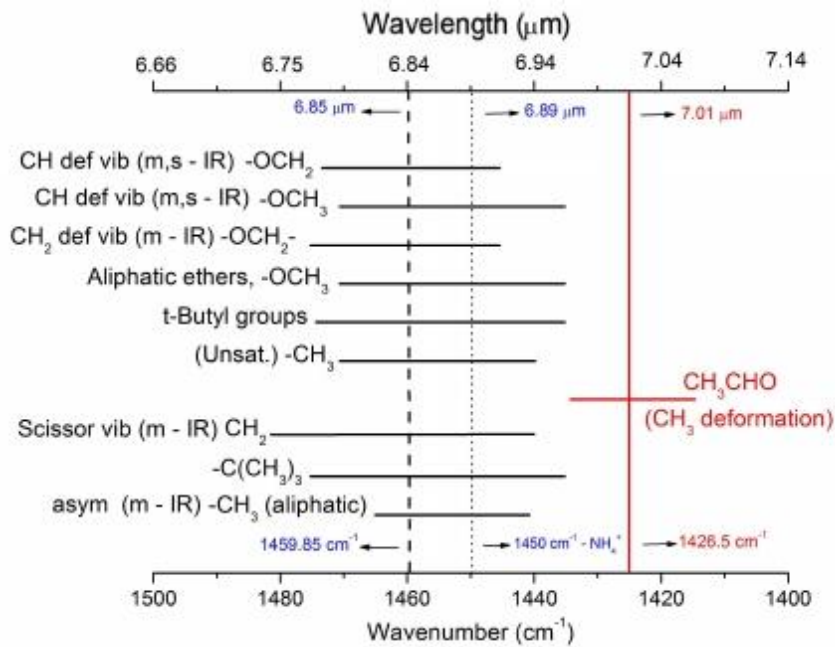

Fig. 10.— Diagram showing molecular species and some radicals containing H, C and O that can be associated with the peak in the IR spectrum within the range of 5.5-8.0 µm . The black dashed line indicates the wavelength at 6.85 µm and the dotted black line indicates the central wavelength associated with the $NH_4$ ion from Schutte & Khanna (2003). The red line indicates the $CH_3$ deformation, associated with acetaldehyde ($CH_3CHO$).

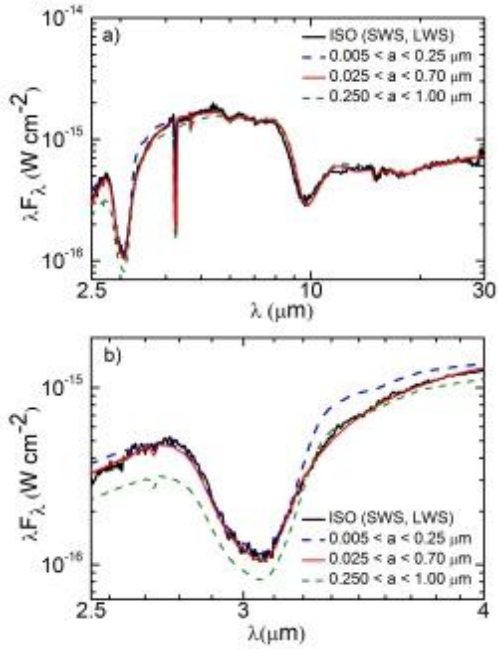

Fig. 11.— Modeling by using different size of grains. Panel a presents the spectrum between 2.5-30 μm, showing that small and large grains cannot to fit the observation. Panel b shows a detail of panel , between 2.5-4.0 μm. The small grains (0.005 - 0.25 μm) can fits small wavelengths, but cannot to fit longer wavelengths. On the other large grains (0.25 - 1.0 μm) cannot to fit the spectrum at near-IR and mid-IR. The better fit is given by intermediary size for grains (0.025 - 0.7 μm).

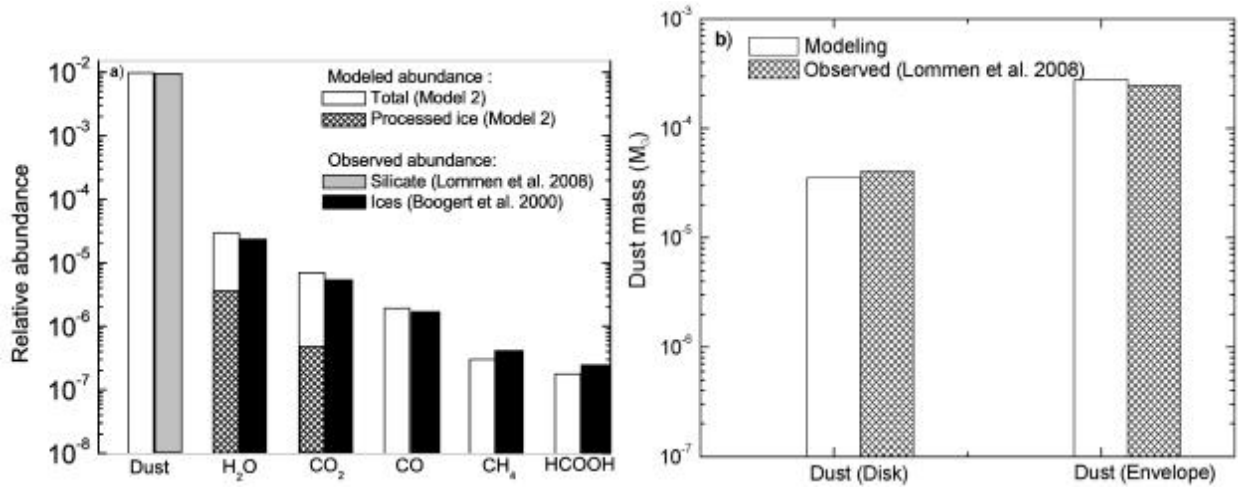

Fig. 12.— a) Relative abundance of each molecular specie used in the modeling compared with observational data from other authors. The white bars represent the total abundance relative to $H_2$, necessary to fit the Elias 29 spectrum, assuming the dust-to-gas ratio equal to 0.01. Likewise, the gray bar is the dust-to-gas ratio taken from Lommen et al. (2008). The hatched bars represent the abundance of the ice processed by ionizing radiation, used in the Model 2. The black bars show the abundances of the ices found in Boogert et al. (2000) for Elias 29 obtained from ratio between column density of the ices and column density of the $H_2$. b) Dust mass (in solar mass) employed in the Model 2 (white bars) for disk and envelope compared with observational data from Lommen et al. (2008) (hatched bars).

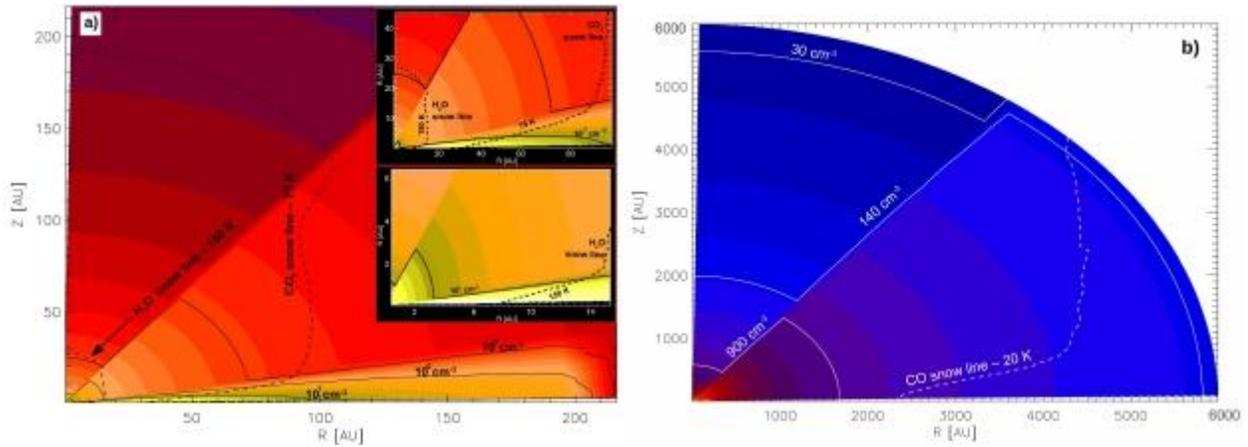

Fig. 13.— Structure of the H2 density (continuous line) in cm$^{-3}$ and selected temperature in K (dashed line) for Elias 29 protostar. Color code indicates changes in the H2 density. a) Region limited to 210 AU to emphasize the disk. The two dashed lines indicate the temperature of 150 K and 75 K, respectively. Inset panels show with details the inner part of the protoplanetary disk. b) Region limited to 6000 AU to emphasize the envelope. The density of that region is very low in relation to the disk, and the dashed line indicates the temperature of 20 K.

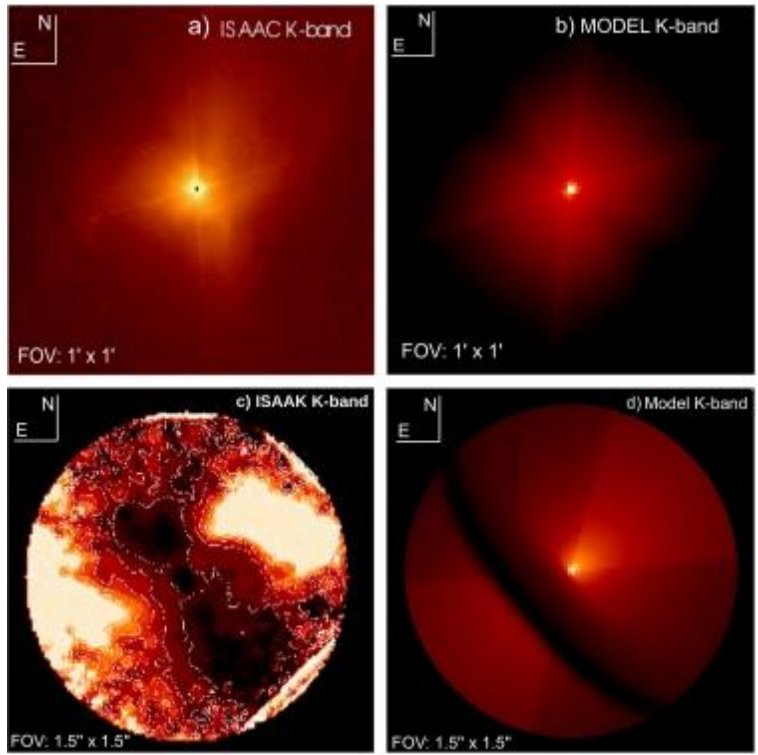

Fig. 14.— Comparison between observed and modeled images of Elias 29 in the infrared with the same field of view (FOV). a) Elias 29 structure in a field of view equal to $1' \times 1'$ on the K-band from ISAAC (Infrared Spectrometer And Array Camera), modified from Hu´elamo et al. (2005). b) Modeled image on the K-band, assuming the parameters of the Model 2. c) Normalized K-band polarization image emphasizing the disk of Elias 29. d) Modeled image calculated from Model 2, without polarization.

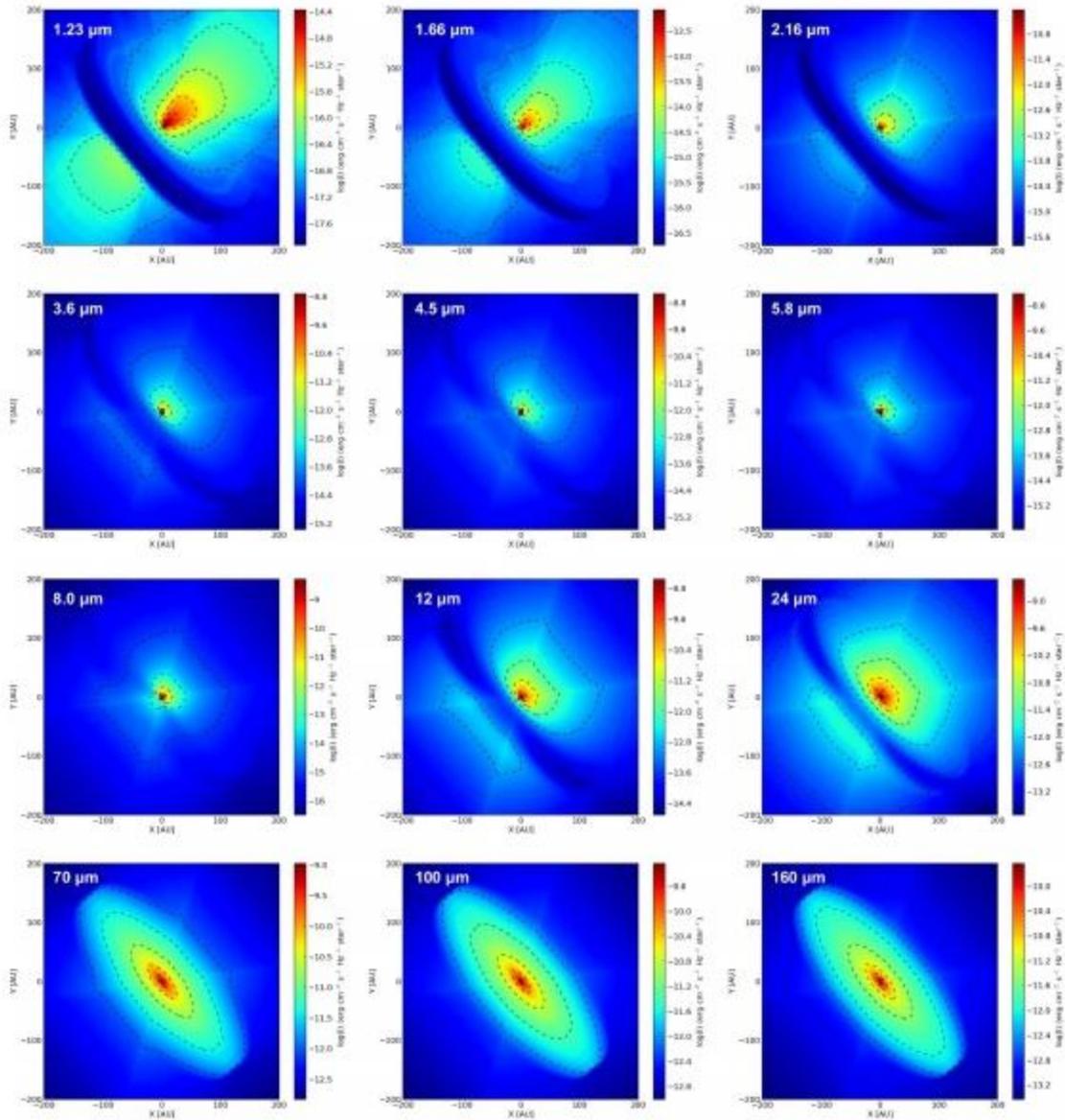

Fig. 15.— Resolved images for Elias 29, modeled using the RADMC-3D code. The layers present the intensities in ergs $cm^{-2}$ $s^{-1}$ $Hz^{-1}$ $sr^{-1}$ for images raging from near to far-IR and the contours correspond to values on the color bar. Each layer has a field of view of $3.3'' \times 3.3''$ or 400 AU $\times$ 400 AU. The inclination of the protoplanetary disk relative to the line of sight is easily observed at low wavelengths (the the upper panels in this figure). An illustration of the modeled scenario is given at Figure 6.

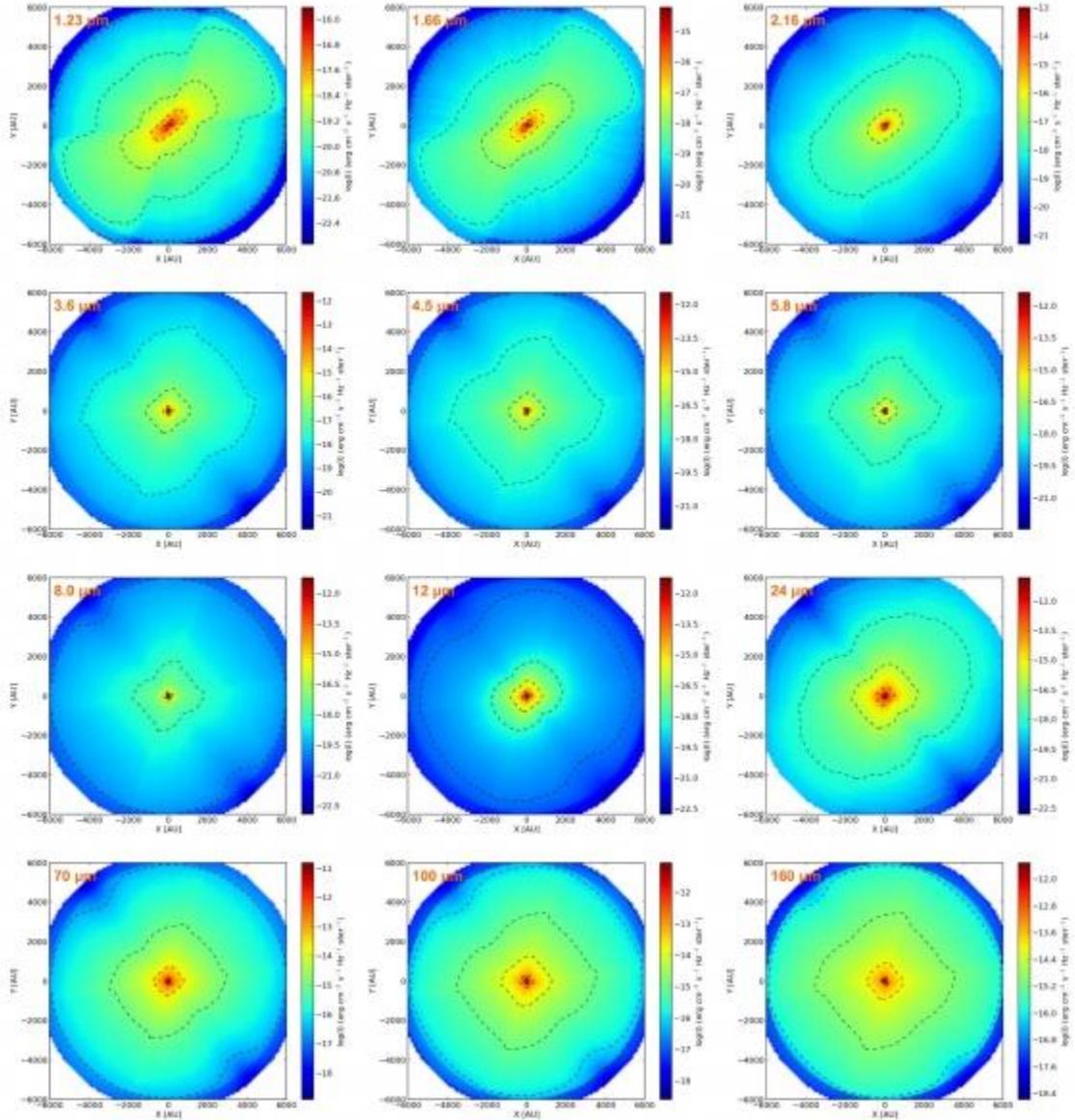

Fig. 16.— Resolved images for Elias 29, modeled using the RADMC-3D code. The layers present the intensities in ergs cm$^{-2}$ s$^{-1}$ Hz$^{-1}$ sr$^{-1}$ for images raging from near to far-IR and the contours correspond to values on the color bar. Each layer has a field of view of 100$''$ × 100$''$ or 12000 AU × 12000 AU.

Table 1: Better parameters employed in the Model 1 (non-bombarded ices) and Model 2 (bombarded ices).

| Parameter | Description | Employed value (see text) | Estimated range | Literature value |
|---|---|---|---|---|
| R ($R_\odot$) | Stellar radius | 5.7 | 5 - 12 | 5.8[a] , 5.9[b] |
| T (K) | Stellar temperature | 4880 | 4000 - 5000 | 4786[b] , 4913[a] |
| L ($L_\odot$) | Stellar luminosity | 16.5 | 13 - 36 | 13.6[c] , 16.3[b] , 17.5[a] 36[d] |
| $M_d$ ($M_\odot$) | Disk mass | 0.003 | 0.002 - 0.007 | < 0.007[e] |
| $R_{d,in}$ (AU) | Disk inner radius | 0.36 | fixed | 0.25[f] |
| $R_{d,out}$ (AU) | Disk outer radius | 200 | fixed | 200[e] |
| $M_{env}$ ($M_\odot$) | Envelope mass | 0.028 | 0.02 - 0.06 | <0.058[e] |
| $R_{env,in}$ (AU) | Envelope inner radius | 0.36 | fixed | - |
| $R_{env,out}$ (AU) | Envelope outer radius | 6000 | fixed | 6000[g] |
| $\theta_c$ ($^\circ$) | Cavity angle | 30 | 25 - 55 | 40[h] |
| d (pc) | Distance | 120 | 100-160 | 125[i] , 160[g] |

Table 2: Parameters obtained from graybody fitting relative to cold foreground clouds.

| Graybody | $N_H$ $(cm^{-2})$[a] | $A_V$ $(mag)$[b] | $\Delta\Omega$ $(sr)$ | $\beta$ | T (K) |
|---|---|---|---|---|---|
| 1 | $(8.0 \pm 0.2) \times 10^{21}$ | $4.0 \pm 1.0$ | $(1.7 \pm 0.1) \times 10^{-8}$ | 2 | $25.2 \pm 1.3$ |
| 2 | $(1.2 \pm 0.3) \times 10^{22}$ | $6.0 \pm 1.5$ | $(7.0 \pm 0.1) \times 10^{-7}$ | 2 | $20.2 \pm 0.8$ |

---

[a] Assuming dust-to-gas ratio as 0.01

[b] Using the relation $N_H = 2 \times 10^{21} A_V$ $cm^{-2}$ $mag^{-1}$ from Bohlin et al. (1978)

Table 3: Model 2 comparison between the observed and modeled $\lambda F_\lambda$ for each absorption band of ice and silicate. The central wavelength and wavenumber, as well as, the associated molecules are presented in the columns 1, 2 and 3, respectively. The $\lambda F_\lambda$ values are presented in the columns 4 and 5, and the percentage error in the column 6.

| $\lambda$ ($\mu$m) | k (cm$^{-1}$) | Assignment | $\lambda F_\lambda$ (Observed) (W cm$^{-2}$)[a] | $\lambda F_\lambda$ (Modeled) (W cm$^{-2}$)[b] | Error (%) |
|---|---|---|---|---|---|
| 2.69 | 3708 | $CO_2$ | $(5.92 \pm 0.05) \times 10^{-16}$ | $5.91 \times 10^{-16}$ | 0.17 |
| 3.07 | 3250 | $H_2O$ | $(1.38 \pm 0.03) \times 10^{-16}$ | $1.37 \times 10^{-16}$ | 0.72 |
| 4.26 | 2342 | $CO_2$ | $(1.36 \pm 0.03) \times 10^{-16}$ | $1.38 \times 10^{-16}$ | 1.45 |
| 4.67 | 2139 | CO | $(1.22 \pm 0.02) \times 10^{-15}$ | $1.21 \times 10^{-15}$ | 0.81 |
| 5.83 | 1715 | HCOOH | $(1.47 \pm 0.03) \times 10^{-15}$ | $1.44 \times 10^{-15}$ | 2.01 |
| 6.00 | 1650 | $H_2O$ | $(1.45 \pm 0.01) \times 10^{-15}$ | $1.44 \times 10^{-15}$ | 0.70 |
| 6.85 | 1459 | $CH_3$ def? | $(1.48 \pm 0.08) \times 10^{-15}$ | $1.47 \times 10^{-15}$ | 0.67 |
| 7.01 | 1426 | $CH_3CHO$? | $(1.40 \pm 0.01) \times 10^{-15}$ | $1.42 \times 10^{-15}$ | 1.41 |
| 7.67 | 1304 | $CH_4$ | $(1.38 \pm 0.02) \times 10^{-15}$ | $1.41 \times 10^{-15}$ | 2.27 |
| 9.70 | 1031 | Silicate | $(2.94 \pm 0.01) \times 10^{-15}$ | $3.00 \times 10^{-15}$ | 2.00 |
| 12.5 | 800 | $H_2O$ | $(5.55 \pm 0.41) \times 10^{-16}$ | $5.70 \times 10^{-16}$ | 2.63 |
| 15.15 | 660 | $CO_2$ | $(4.53 \pm 0.29) \times 10^{-16}$ | $4.52 \times 10^{-16}$ | 0.22 |
| 15.25 | 655 | $CO_2$ | $(4.78 \pm 0.28) \times 10^{-16}$ | $4.87 \times 10^{-16}$ | 1.84 |
| 18.00 | 555 | Silicate | $(5.27 \pm 0.17) \times 10^{-16}$ | $5.20 \times 10^{-16}$ | 1.30 |

[a] ISO (SWS, LWS) de Graauw et al. (1996) and Clegg et al. (1996)

[b] This paper

Table 4: Average abundances of ices observed toward Elias 29 calculated from Model 2, compared with values obtained in the literature.

| Ice species | Column density ($10^{18}$ cm$^{-2}$) | Abundance relative to H$_2$ gas | Abundance relative to H$_2$ gas (Literature)[a] | Wavelength - μm (Wavenumber - cm$^{-1}$) | Band strength cm molecule$^{-1}$ |
|---|---|---|---|---|---|
| H$_2$O | 3.50 | $2.9 \times 10^{-5}$ | $2.5 \times 10^{-5}$ | 3.07 (3250) | $2.0 \times 10^{-16}$[b] |
| CO$_2$ | 0.84 | $7.0 \times 10^{-6}$ | $5.6 \times 10^{-6}$ | 4.26 (2342) | $7.6 \times 10^{-17}$[b] |
| CO | 0.21 | $1.9 \times 10^{-6}$ | $1.7 \times 10^{-6}$ | 4.67 (2139) | $1.1 \times 10^{-17}$[b] |
| HCOOH | 0.02 | $1.8 \times 10^{-7}$ | $< 2.5 \times 10^{-7}$ | 5.83 (1715) | $6.7 \times 10^{-17}$[c] |
| CH$_4$ | 0.04 | $3.1 \times 10^{-7}$ | $< 4.1 \times 10^{-7}$ | 7.67 (1304) | $7.3 \times 10^{-18}$[d] |
| CH$_3$OH | 0.06 | $1.3 \times 10^{-7}$ | $< 6.3 \times 10^{-7}$ | 6.85 (1460) | $1.2 \times 10^{-17}$[e] |
| CH$_3$CHO | 0.004 | $3.2 \times 10^{-8}$ | - | 7.01 (1426) | $3.6 \times 10^{-18}$[f] |

[a] Boogert et al. (2000)

[b] Gerakines et al. (1995)

[c] Shutte et al. (1999)

[d] Kerkhof et al. (1999)

[e] Assuming to be the CH$_3$ deformation of CH$_3$OH at 6.85 μm

[f] Bennett et al. (2005) - Assuming to be the CH$_3$CHO molecule